\documentstyle[epsfig,tighten,preprint,aps]{revtex}

\begin{document}
\draft

\title{On the Migdal-Watson approach to 
FSI effects in meson production in $NN$ collisions}

\author{V. Baru$^{a,b}$, A.M. Gasparian$^{a,b}$, J. Haidenbauer$^{a}$, 
A.E. Kudryavtsev$^{b}$, and J. Speth$^{a}$}

\address{
$^a$Institut f\"{u}r Kernphysik, Forschungszentrum J\"{u}lich
GmbH, D--52425 J\"{u}lich, Germany \\
$^b$Institute of Theoretical and Experimental Physics, 
117258, B. Cheremushkinskaya 25, \\
Moscow, Russia }

\maketitle 

\begin{abstract}

The influence of the nucleon-nucleon final state interaction
(FSI) on properties of the meson production amplitude near threshold is
discussed. For the nucleon-nucleon interaction a simple Yamaguchi potential
as well as realistic potential models are considered.
It is shown that FSI effects cannot be factorized from the
production amplitude. The absolute magnitude of FSI effects depends on the 
momentum transfer (or on the mass of the produced meson) and hence
is not universal. Only in the case of the production of rather heavy mesons 
like $\eta^{\prime}$ or $\phi$ FSI effects become universal.
The Jost function approach to FSI effects is critically examined. 

\vskip 0.5cm

\end{abstract}
\newpage

\section{Introduction}

Already in the 1950's 
K. Watson \cite{Wat} and A. Migdal \cite{Mig} have shown that
the energy dependence for meson production reactions $NN\to NNx$
near threshold is predominantly determined by the strong $NN$ 
interaction in the final state. Their arguments have been used 
for justifying a rather simple treatment of effects from the
final state interaction (FSI) (see, e.g., Refs. \cite{Moa,Ged,Ber,Kai}). 
It consists in simply multiplying the basic production amplitude with 
the on-shell $NN$ $T$-matrix, i. e.  
\begin{equation}
{\cal M}=- N A_{prod}^{on}\cdot\frac{e^{i\delta}\sin\delta}{ka_{NN}},
\label{Mtot}
\end{equation}
where $\delta=\delta(k)$ is the $NN$ phase shift, $a_{NN}$ the $NN$ 
scattering length, $A_{prod}^{on}$ the on-shell meson production 
amplitude and $N$ a normalization factor. This expression suggests
that the FSI effect is universal, i. e. does not depend on the
specific meson emitted.

Recently, some aspects of FSI effects in the reaction $NN\rightarrow NNx$ 
were investigated by Hanhart and Nakayama \cite{Han} and Niskanen \cite{Nis}. 
In particular, these authors pointed out that
the evaluation of the total reaction amplitude by just multiplying the
production amplitude by the on-shell $NN$~$T$-matrix is not acceptable
for obtaining quantitative predictions. In the present paper we 
want to study those FSI effects in more detail. Specifically, we
want to shed some light on the validity of the multiplication prescription 
Eq. (\ref{Mtot}). We examine the influence of the $NN$ FSI 
on the absolute value of the 
reaction amplitude by employing realistic models of the
$NN$ interaction. Furthermore we investigate
the dependence of the FSI effects on the mass of the produced
meson. For that purpose we will vary the mass of $x$ and adopt values 
corresponding to those of the $\pi$, $\eta$, and $\eta^{\prime}$ mesons.

\begin{center}
\begin{figure}
\epsfig{file=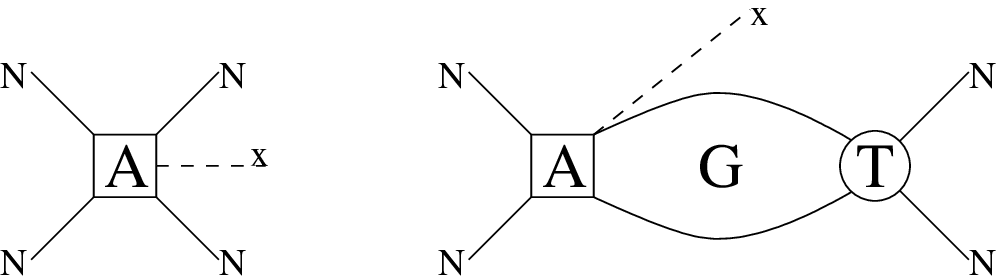, width=12.0cm}
\vspace*{1cm}
\caption{
Diagrammatic representation of the DWBA expression Eq. (\ref{Mgen}). 
$A$ is the elementary meson-production amplitude and $T$ the $NN$ 
T-matrix.}
\label{fig1}
\end{figure}
\end{center}

In general the total amplitude for the reaction $pp\rightarrow ppx$ can 
be determined from the DWBA expression
\begin{equation}
{\cal M}=A_{prod}^{on}+ A_{prod}^{off}G_0 T_{NN}.
\label{Mgen}
\end{equation}
where the term on the very right side implies an integration over the
off-shell production amplitude and the off-shell $NN$ $T$-matrix. 
Eq.~(\ref{Mgen}) corresponds to the sum of the two
diagrams shown in Fig.~\ref{fig1}. 
Meson production in $NN$ collisions requires a large momentum
transfer between the initial and final nucleons which is typically of the 
order of $\sqrt{mm_x}$, where $m$ is the nucleon mass and $m_x$ the mass of 
produced meson. Thus the range of the production interaction
will be much smaller than the characteristic range of the $NN$ interaction 
in the final state. Goldberger and Watson argued that in such a case 
the meson can be considered to be produced practically from a point like 
region so that the production amplitude can be factored out of the
integral \cite{Gol}, i.e. 
\begin{equation}
{\cal M}=A_{prod}^{on}+ A_{prod}^{off}G_0 T_{NN} \ \approx \ 
A_{prod}^{on} \lbrack 1 + G_0 T_{NN} \rbrack \ = \ 
A_{prod}^{on}{\psi_k^{(-)}}^{*}(0).
\label{MDru}
\end{equation}
Here $\psi_k^{(-)}(\vec{r})$ is the (suitably normalized) $NN$
wave function in the continuum \cite{Gol}, where  
$\psi_k^{(-)}(0)$ is related to the Jost function 
${\cal J}$ via $ {\psi_k^{(-)}}^{*}(0) \ = {\cal J}^{-1}(-k)$ 
\cite{Gol}.

Clearly also in this case one arrives at results where the FSI effects
are reduced to a mere multiplicative factor
$ \mid {\psi_k^{(-)}}^{*}(0) \mid^2$ (commonly referred to as
enhancement factor). 

The prescription Eq. (\ref{MDru}) has been utilized by several authors
\cite{Dru,Wil,Shy,Sib,Tit} in their studies of meson production. 
Its validity has been examined by an explicit calculation of the loop 
diagram in Ref.~\cite{Dru} employing an $OBE$ model for the production 
amplitude. However, one has to keep in mind that this investigation 
is based on a simple separable Yamaguchi potential for the $NN$ FSI. 
It is well-known that the off-shell behavior of the $T$-matrix for 
the Yamaguchi potential is rather different from the one
resulting from realistic models of the $NN$ force.
This can be seen from Fig.~\ref{fig2} where we compare the
off-shell properties of the Paris \cite{Lac} and (one version of)
the Bonn \cite{Bonn} $NN$ models with the one
of the Yamaguchi potential for the $^1S_0$ partial wave. 
The most striking difference is definitely the zero crossing of the 
$T$-matrix that occurs for realistic potential models 
at off-shell momenta $q\approx 350$ MeV/c whereas the one of the
Yamaguchi potential never changes sign. As we will show below, 
this specific feature has a strong and important influence on the result 
for the FSI effects.

The paper is structured in the following way. In Sect. 2 we 
present our formalism. We specify the meson production amplitude
that we use in the present investigation and we give the explicit
expression for the loop diagram of Fig.~\ref{fig1}b. In Sect. 3
we present and discuss our results. Specifically, we show calculations 
for the effects of the FSI considering different $NN$ models and
the production of mesons with different masses. Furthermore we
take a look at the energy dependence of the FSI effects
and examine the validity of some commonly used approximations. 
The paper ends with a short summary. 

\section{Loop diagram calculus}

For the calculation of the loop diagram of Fig.~\ref{fig1}b with 
off-shell amplitudes of realistic $NN$ interactions we need to specify 
a model for the production amplitude. We assume that it has the form
\begin{equation}
A_{prod}=\frac{g \cdot A_{\mu N\to xN}}{t-m_\mu^2} \ , 
\label{5}
\end{equation}
which corresponds to the exchange of a scalar meson $\mu$ of mass $m_\mu$ in
the $t$-channel followed by the production of a meson $x$ in 
a rescattering process. 
The corresponding diagram is shown in Fig.~\ref{fig3}a. 
The coupling $g$ at the $NN\mu$ vertex and the amplitude $A_{\mu N\to xN}$ 
are assumed to be constants. For $m_\mu$ we take the value of the pion
mass, i.e. $m_\mu$ = 135 MeV. Furthermore, for simplicity reasons, we 
use non-relativistic kinematics for the intermediate nucleons. 
The total reaction amplitude for this production model is then given
by the sum of the two diagrams of Fig.~\ref{fig3}, i.e. 
\begin{equation}
{\cal M}=-\frac{mg}{E}
\frac{A_{\mu N\to x N}}{[(\vec k -\vec P/2 +\frac{m}{E}\vec p)^2+\lambda^2]}
\Psi (\vec k) \ ,
\label{summ}
\end{equation}
where $E=\sqrt{m^2+p^2}$, $\lambda^2=\frac{m}{E}m_\mu^2+\frac{m^2}{E^2}\tau^2$,
with $\tau=E-m$. $\Psi (\vec k)$ is given by the expression
\begin{equation}
\Psi (\vec k)=1-
\frac{m\pi [(\vec k-\vec P/2+\frac{m}{E}\vec p)^2+\lambda^2]}{r} 
\int^{\infty}_0\frac{dqq \ T_{NN}(q,k)}{[q^2-k^2-i0]} \ 
ln[\frac{(q+r)^2+\lambda^2}{(q-r)^2+\lambda^2}] \ ,
\label{summ1}
\end{equation}
where $r=\mid\frac{m}{E}\vec p-\vec P/2\mid$.
$T_{NN}(q,k)$ is the $NN$ half-off-shell T-matrix in the $^1S_0$
partial wave. 
The function $F_{NN}(k)=\mid\Psi (\vec k)\mid^2$ can
be considered as a generalization of the FSI enhancement factor 
$\mid{\psi^{(-)}}^{*}_k(0)\mid^2$ that follows from the 
factorization assumption Eq. (\ref{MDru}). We would like to
emphasize, however, that (contrary to 
${\psi^{(-)}}^{*}_k(0)$ in Eq. (\ref{MDru})) 
$\Psi (\vec k)$ does contain also information on the production mechanism 
and not only on the $NN$ FSI.

In the actual calculations we want to include the Coulomb interaction between
the outgoing protons. Therefore we have to replace the $NN$ half-off-shell
$T$-matrix in Eq. (\ref{summ1}) by the quantity $T^{cs}_{NN}$, i.e. the
Coulomb-distorted hadronic $T$-matrix. This quantity is obtained by the
prescription introduced in Ref.~\cite{Han0}, namely via
\begin{equation}
T^{cs}_{NN}(q,k) = \frac{C(\gamma_q)}{C(\gamma_k)} 
\frac{T_{NN}(q,k)}{T_{NN}(k,k)} T^{cs}_{NN}(k,k) \ , 
\label{Coul}
\end{equation}
where $k$ and $q$ denote the on-shell and off-shell momentum, respectively.
$T_{NN}(k,k)$ and $T_{NN}(q,k)$ are the on-shell and half-off-shell 
$T$-matrices for the strong interaction alone. The Coulomb penetration 
factor $C$ is given by
\begin{equation}
C^2(\gamma_q) = \frac{2 \pi \gamma_k}{e^{2\pi\gamma_k}-1}; \ \ 
\gamma_k = \frac{m}{2}\frac{1}{\alpha k} \ , 
\label{Pene}
\end{equation}
with $\alpha$ the fine-structure constant. 
Furthermore, the first term on the l.h.s. of Eq. 
(\ref{summ1}) (the ``1'') has to be replaced by $C(\gamma_k)$.

\section{Discussion}

First we want to discuss the dependence of $\Psi (k)$ on the mass of the 
produced meson. For that purpose we start out from a somewhat simpler 
expression for $\Psi $ which follows from Eq. (\ref{summ1}) 
for the kinematics at the production threshold:
\begin{equation}
\Psi (k)=C(\gamma_k)-
\frac{m\pi [mm_x+m_\mu^2]}{\sqrt{mm_x+m_x^2/4}}
\int^{\infty}_0\frac{dqq \ T^{cs}_{NN}(q,k)}{[q^2-k^2-i0]} \ 
ln[\frac{(q+\tilde r)^2+\tilde\lambda^2}{(q-\tilde r)^2+\tilde\lambda^2}] \ ,
\label{summ2}
\end{equation}
where
$$\tilde r=\frac{m}{m+m_x/2}\sqrt{mm_x+m_x^2/4}, \
\tilde\lambda^2=\frac{m^2}{(m+m_x/2)^2}\frac{m_x^2}{4}+
\frac{m}{m+m_x/2}m_\mu^2.$$

For the production of a light meson, $m_x\ll m$, we get 
$\tilde r\approx \sqrt{mm_x}$, $\tilde \lambda^2=m_\mu^2+m_x^2/4$, 
so that there is a dependence of the integral on the r.h.s. of 
Eq. (\ref{summ2}) on $m_x$.
In the case of a heavy meson, $m_x\gg m$, it follows that 
$\tilde r\approx m$, $\tilde\lambda^2\approx m^2$ and consequently
$\Psi (k)$ does not depend on the mass of the emitted
meson $x$. In other words, we expect that FSI effects are getting 
universal for the production of heavy mesons 
via an OBE-type production mechanism Eq.~(\ref{5}). 

\begin{center}
\begin{figure}
\epsfig{file=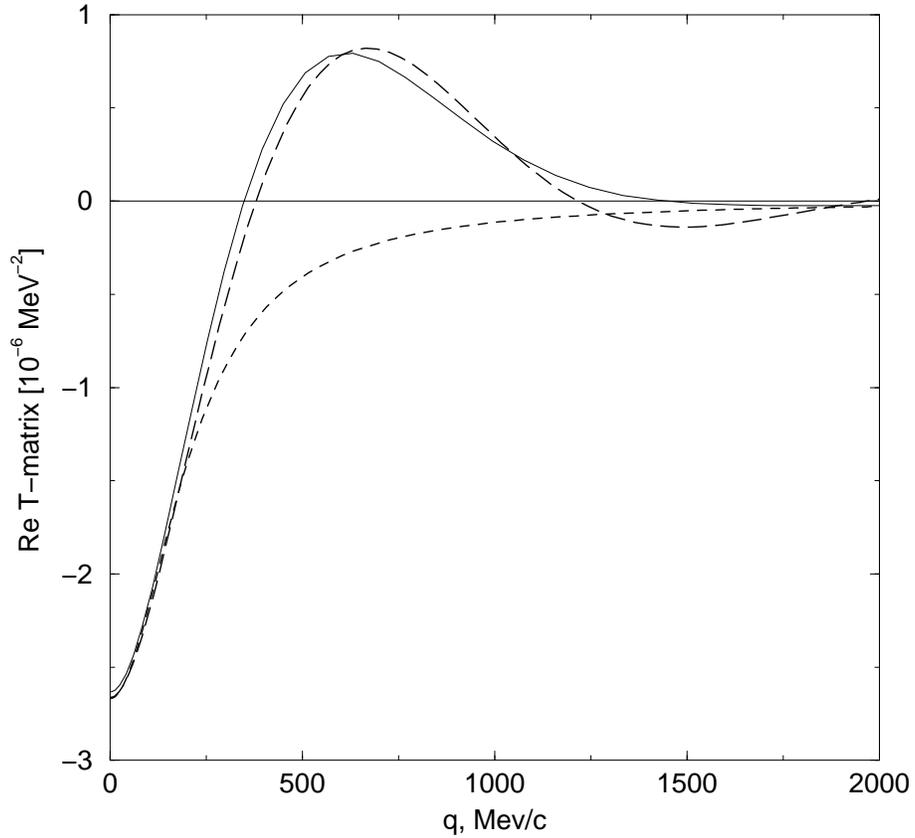, width=12.0cm}
\caption{
Real part of the $NN$ $^1S_0$ T-matrix as a function of the off-shell 
momentum $q$ calculated at the fixed on-shell momentum $k$ = 10 MeV/c.
The solid, long-dashed, and short-dashed lines are the results for the
Paris \protect\cite{Lac}, Bonn \protect\cite{Bonn}, 
and Yamaguchi potentials, respectively.}
\label{fig2}
\end{figure}
\end{center}

\begin{center}
\begin{figure}
\epsfig{file=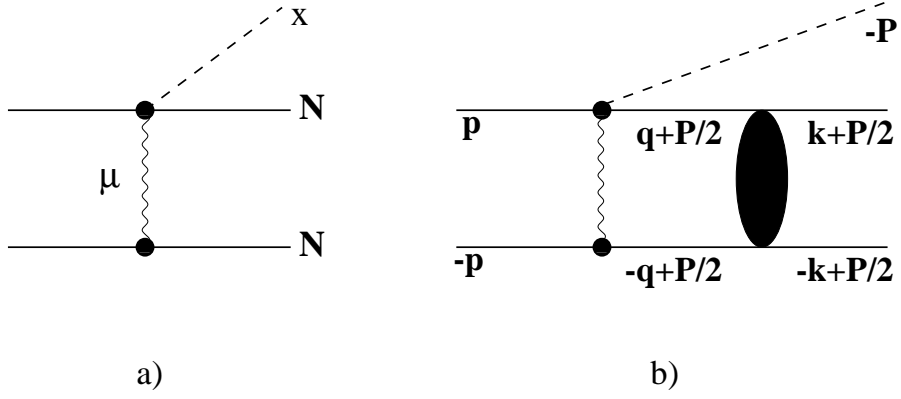, width=12.0cm}
\vspace*{1cm}
\caption{Contributions to the total reaction amplitude ${\cal M}$:
(a) Born term $A$; (b) loop diagram including the final state interaction.
}
\label{fig3}
\end{figure}
\end{center}

\begin{figure} [t]                                      
\vskip 9.5cm    
\includegraphics{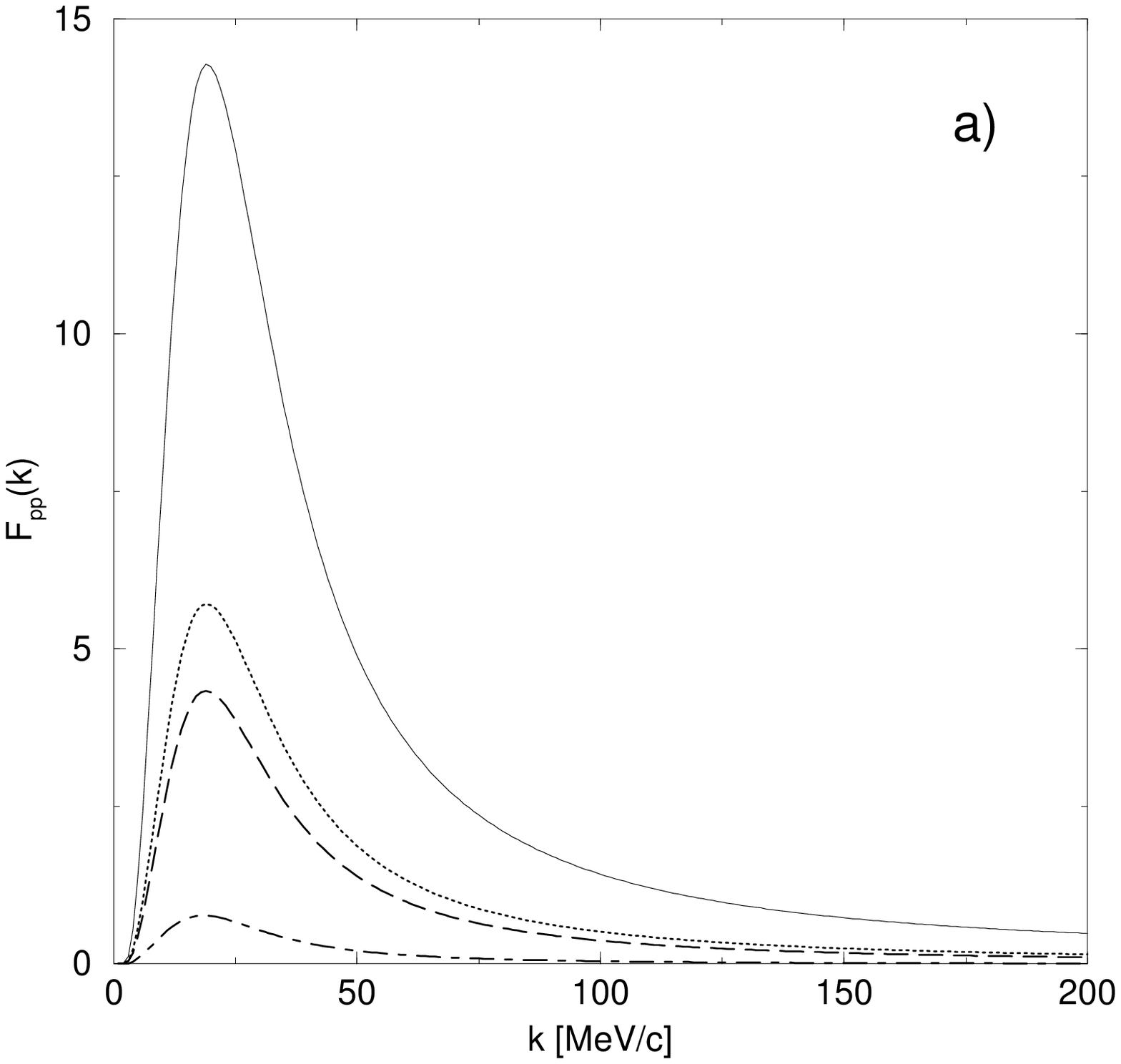}
\includegraphics{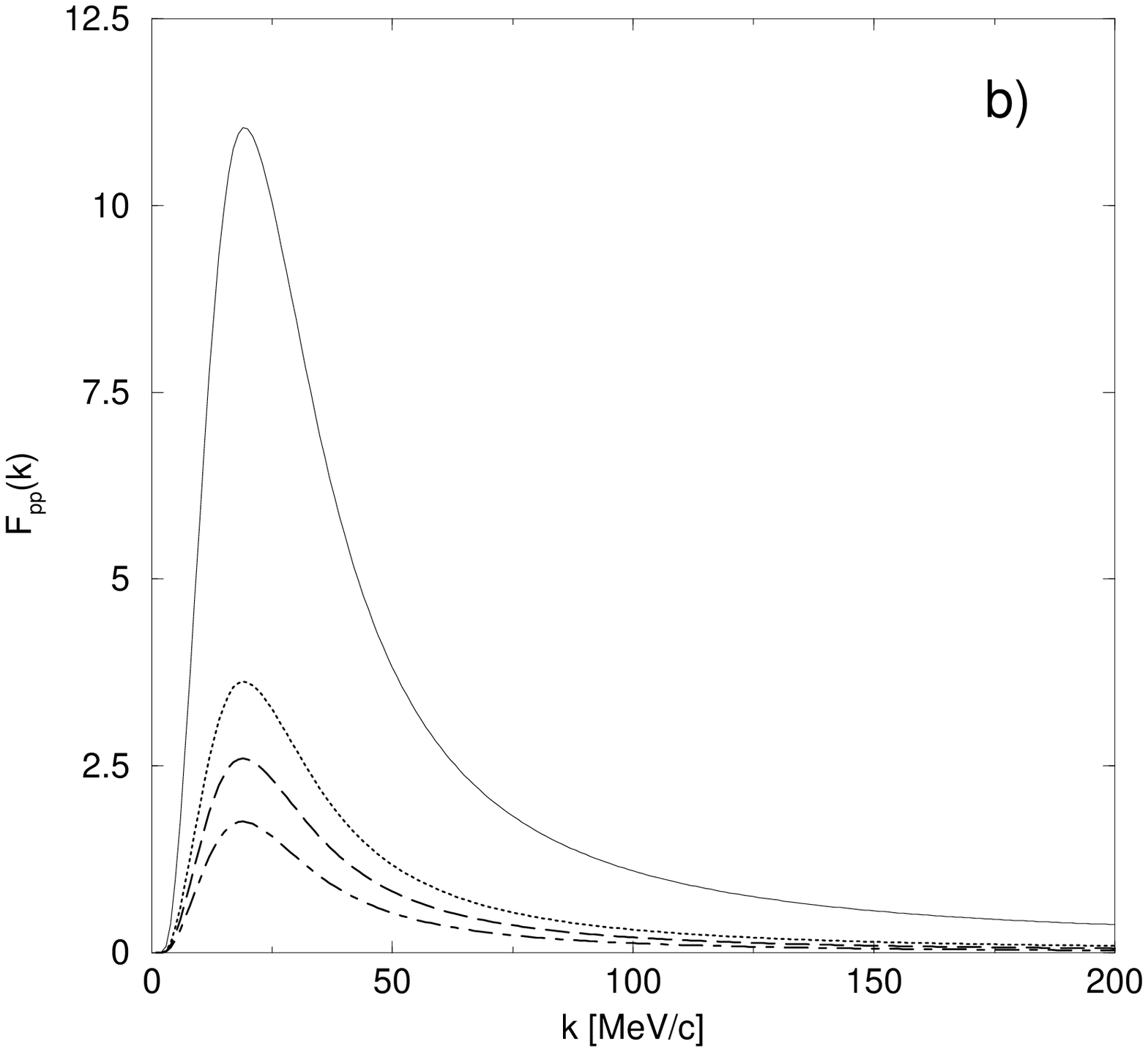}
\includegraphics{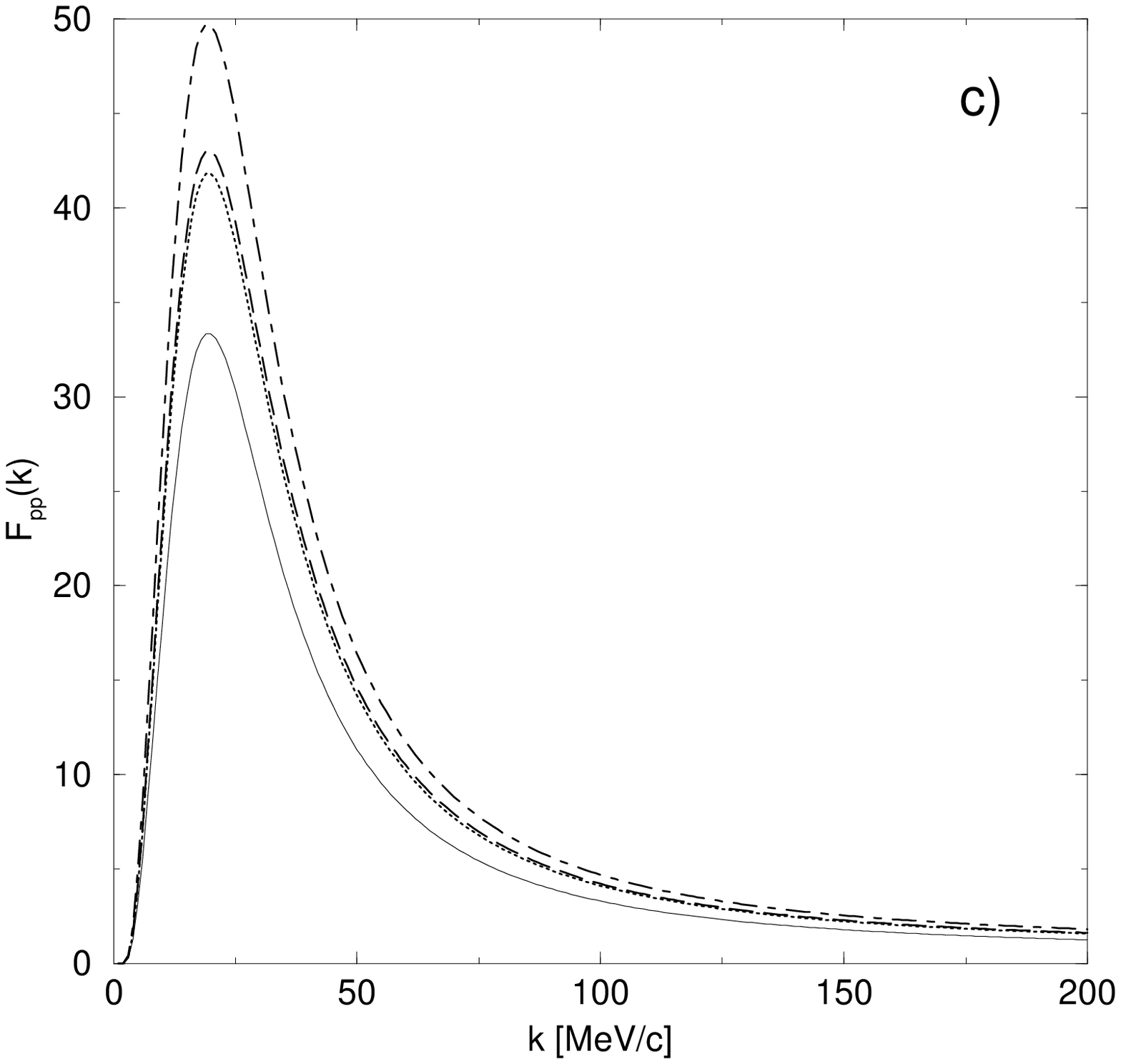}
\vspace*{20.0cm}
\caption{The FSI factor $F_{pp} \ = \ \mid \Psi (k) \mid^2$ 
(cf. Eq. (\ref{summ2}))    
for the Paris (a), Bonn (b), and the Yamaguchi (c) potentials 
and the production of the $\pi$ (solid curve), $\eta$ (dotted curve),
and $\eta '$ (dashed curve) mesons. The dash-dotted lines are the 
results based on the factorization assumption Eq. (\ref{MDru}), i.e. 
$F_{pp}(k) \ = \ \mid {\cal J}(k) \mid^{-2}$.  
Note that different scales are used for each $NN$ model!} 
\label{fig4}
\end{figure}

Let us now come to the results for the FSI factor
$F_{NN}=\mid \Psi (k)\mid^2$. In Fig.~\ref{fig4} we show calculations 
for different $NN$ models and for some typical masses of the 
emitted meson $x$.
It can be seen from those figures that the magnitudes of $F_{NN}$
resulting for the Bonn and the Paris potentials are fairly similar 
whereas the one for the separable Yamaguchi potential is quite different.
(Note that different scales are used for each $NN$ model!) 
This result can be understood qualitatively from the features of the
corresponding off-shell T-matrices shown in Fig.~\ref{fig2}. 
The T-matrices for the Bonn and Paris potentials are very similar.
In particular, for both models there is a change of sign at an
off-shell momentum of $q \approx $ 350 MeV/c. Because of this change of 
sign cancellations occur in the integral for $\Psi (k)$
(cf. Eq.~\ref{summ1}). The off-shell T-matrix of the Yamaguchi 
potential does not change sign. Therefore, no such cancellations take place
in the integration and as a consequence the FSI factor $F_{NN}$ is 
significantly larger than the ones for the realistic interaction models, 
cf. Fig.~\ref{fig4}. 

\begin{figure}
\hspace*{0.6cm}\epsfig{file=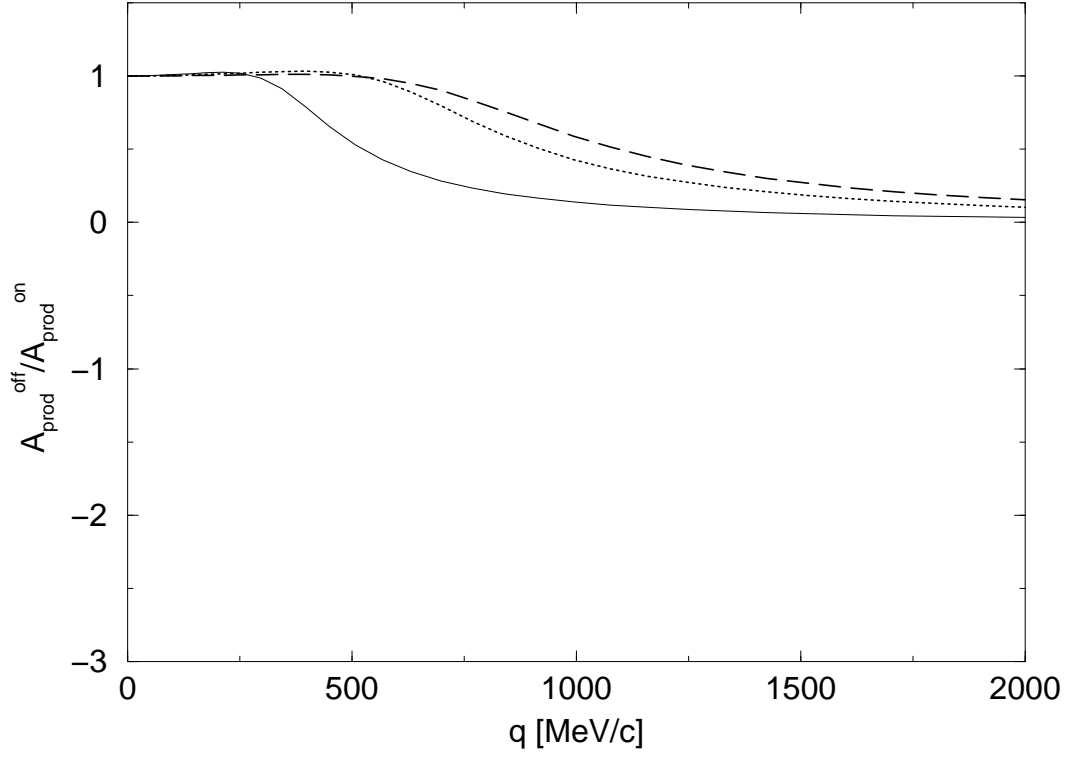, width=14.0cm}
\caption{
Ratio $A_{prod}^{off}/A_{prod}^{on}$ of the production amplitude 
as a function of the off-shell momentum $q$ calculated at the fixed 
on-shell momentum $k$ = 0 MeV/c. The solid, dotted and dashed
curves correspond to the production of the $\pi$, $\eta$, and
$\eta '$ mesons, respectively. 
} 
\label{fig5}
\end{figure}

There is also a striking difference in the
results with regard to the mass of the emitted meson. For the Paris
and Bonn potentials the FSI factor decreases with increasing
mass of the produced meson. However, for the Yamaguchi potential we
observe the opposite effect. Here $F_{NN}(k)$ becomes larger with the mass
of the produced meson. These features can again be understood in terms
of the $NN$ off-shell properties. However, now the off-shell behavior 
of the production amplitude, that enters into the
integral (\ref{summ1}) as well, becomes also relevant. 
With increasing mass of the produced meson the required momentum transfer 
$t$ increases as well and, accordingly, the production mechanism gets more 
and more short-ranged. As a consequence the production amplitude remains
a constant over a larger (off-shell) momentum range, as can be seen in
Fig.~\ref{fig5}. This feature enhances the cancellation effects for
the Bonn and Paris $NN$ T-matrices, discussed above, and therefore leads
to a reduction of $F_{NN}$ for larger meson masses. In case of the
Yamaguchi potential no such cancellations can occur and therefore the
FSI factor turns out to be almost independent of the mass of 
the produced meson. 

Nevertheless, we see that also for realistic $NN$ potentials 
the FSI factors become more and more similar with increasing 
mass of the produced meson, i.e. for high momentum transfers. This is
expected. It simply reflects the universality of FSI effects for 
the production of heavy mesons as discussed above. 
We would like to emphasize that the universality of FSI effects 
at large $t$ should set in 
not only for the particular production amplitude used in the present
investigation (cf. Eq. (\ref{5})), but is expected to occur in general. 
Actually, we examined the behavior of $F_{NN}$ for the OBE-type 
production amplitude Eq. (\ref{5}) with inclusion of form factors 
of monopole and dipole type at the $NN\mu$ vertex. Corresponding 
numerical calculations clearly indicate that the qualitative behaviour 
of the FSI factors remains basically unchanged. 

However, it is important to realize that the actual values for
the FSI factors do, of course, depend on the particular production 
amplitude. Thus, the results presented in Figs.~\ref{fig4} are {\it by not
means} absolute predictions that can be taken from this paper and used 
blindly for FSI corrections in any other study of meson production. Rather 
our results demonstrate that FSI effects have to be calculated
always explicitly, utilizing the respective production amplitudes and a 
proper $NN$ off-shell T-matrix, if one wants to obtain reliable
quantitative predictions. (In this context we also would like to draw 
attention to the requirement of a consistent treatment of both the $NN$
scattering and production amplitudes as discussed in the Appendix
of Ref.~\cite{Han1}.) 
Specifically, this means that the apparent 
universality of the FSI effects for large meson masses {\it does not}
imply that one can use the prescription applied in the studies 
\cite{Dru,Shy,Sib,Tit}, i.e.
take the production amplitude out of the integral in Eq. (\ref{summ2}). 
Even though the factor $\Psi (k)$ becomes independent of the mass of
the produced meson its actual magnitude is still determined by the
off-shell properties of the $NN$ T-matrix as well as by the production
amplitude. In order to demonstrate this we show also results based on
the factorization assumption Eq. (\ref{MDru}) (dash-dotted curves). 
In this case the FSI factor is simply given by the enhancement factor
$F_{NN}(k) \ = \ \mid {\cal J}(k) \mid^{-2}$. It is really startling 
how strongly the results for the Yamaguchi potential and for realistic
$NN$ interaction models differ. For the former potential the
enhancement factor based on the Jost function is larger than the
FSI factor obtained from Eq. (\ref{Mgen}) whereas for the latter
models it turns out to be much smaller than the DWBA values. Clearly,
these results suggest that it is rather questionable to use the Jost
function of some arbitrary potentials for the evaluation of FSI 
effects in meson-production reactions \cite{Dru,Shy,Sib,Tit}.

Finally a remark on the differences between the results for 
realistic $NN$ models. Obviously these are small (about 20 \%) for
the pion-production case. But for the $\eta '$ meson the Paris
result is almost a factor 2 larger then the one for the Bonn model.
This is not too surprising because for the production of heavier mesons
a larger momentum transfer is required and therefore the features of
the $NN$ interaction at shorter distances (or larger off-shell momenta)
become more and more important in the actual calculations. As we can
see in Fig.~\ref{fig2} there are fairly large differences in the
off-shell properties of these two models for large off-shell momenta. 
Note that such a sensitivity to the off-shell behaviour of realistic
$NN$ models at large off-shell momenta has been also seen, e.g., in
proton-proton bremsstrahlung producing hard photons \cite{Naka}. 

\begin{figure}
\epsfig{file=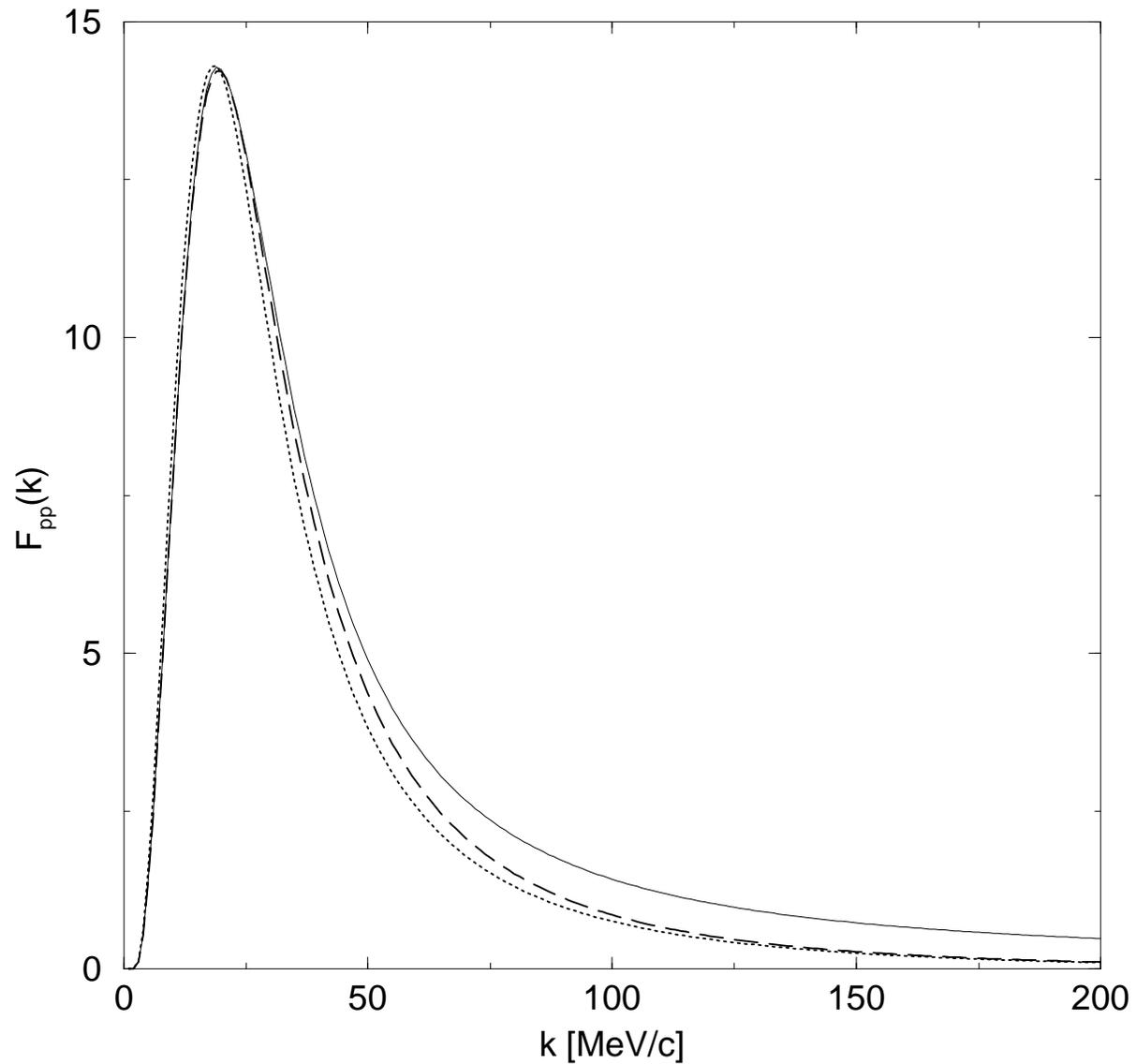, width=16.0cm}
\caption{
The FSI factor $F_{pp} \ = \ \mid \Psi(k) \mid^2$ 
for the Paris $NN$ potential and for pion production 
(solid line) in comparison to results based on the approximations 
Eqs. (\ref{Mtot}) (dashed line) and (\ref{MDru}) (dotted line). The latter
two curves are normalized to the one of the Paris potential at the 
peak ($k \approx$ 20 MeV/c).}
\label{fig6}
\end{figure}

By all the variations we see in the magnitudes of the FSI factors 
presented in Fig.~\ref{fig4} we would like to point out that their energy 
dependence is basically the same for all the different potentials and for 
the different meson masses. If we normalize them to the same value 
for small $k$ (e. g., at the peak of $F_{NN}$ at $k \approx$ 20 MeV/c) 
all curves would lie essentially on top of each other. This means that
the energy dependence of the FSI factors is really primarily
determined by the on-shell $NN$ T-matrix. 
Consequently, the on-shell prescription Eq. (\ref{Mtot}) is indeed
a fairly good approximation, at least for energies near the production
threshold. In order to demonstrate this let us compare 
one of the curves based on the Paris potential with the result corresponding 
to Eq. (\ref{Mtot}) (normalized to the Paris curve at $k \approx$ 20 MeV/c), 
cf. Fig.~\ref{fig6}. None-the-less, we do observe an increasing
difference between these two curves for $k \ge$ 50 MeV/c, which corresponds
to excess (cms) energies $Q \ge$ 3 MeV. For $k$ = 100 MeV/c 
($Q \approx$ 10 MeV) the curves differ already by a factor of around 2. 
It is interesting to see that the Jost function
approach (Eq. (\ref{MDru})) deviates even more strongly from the
correct results than the on-shell approximation in the energy range
$k \le$ 100 MeV/c.

Let us now investigate the origin of those discrepancies 
in more detail. For that purpose we re-write the amplitude ${\cal M}$ 
Eq. (\ref{Mgen}) in the form given in Ref. \cite{Han} 
\begin{eqnarray}
{\cal M}&=&-A_{prod}^{on}\frac{e^{i\delta}\sin\delta}{ka_{pp}} \cdot 
[P(k)-a_{pp} \ k \ \cot\delta] \nonumber \\  
&=&-A_{prod}^{on}\frac{e^{i\delta}\sin\delta}{ka_{pp}} \cdot 
[P(k)+1-1/2a_{pp}r_0k^2+O(k^4)+...]
\label{Pk}
\end{eqnarray}
Here $P(k)$ is proportional to the principal value of the loop integral
(cf. Fig.~\ref{fig3}b) and contains the information on the off-shell behavior
of both $A_{prod}$ and $T_{NN}$. (Note that we have neglected corrections
coming from the Coulomb interaction in Eq. (\ref{Pk}) for simplicity
reasons. Those terms don't play a role anymore at the energies were the
discrepancies discussed above occur.)

Evidently corrections to the simple on-shell prescription Eq. (\ref{Mtot}) 
come from the energy dependence of the function $P(k)$ as well as from
the $\cot \delta$ term. Actual calculations with the $NN$ potentials 
utilized in the present study revealed that the value of $P(k)$ at 
$k$ = 0 is positive and about 3 to 5 units large which makes
it to be the dominant piece of the terms in the bracket of Eq. (\ref{Pk}).
Furthermore, $P(k)$ is slowly decreasing with $k$. The $k^2$ term is 
slowly increasing with $k$ 
(Note that $a_{pp}$ is negative for the $^1S_0$ partial wave!) so that 
there is a compensation in the energy dependence of the terms in the 
brackets of Eq.~(\ref{Pk}). This circumstance is certainly partly
responsible for the fact that Eq. (\ref{Mtot}) works relatively well.

As already mentioned above, the value at $P(k=0)$ is positive and fairly
large (cf. also the comments in Ref.~\cite{Han}). 
Note that in order to get Eq. (\ref{Mtot}) with the normalization
$N$ set to one, as chosen in Refs.~\cite{Ber,Kai}, we 
have to assume that $P(k)\equiv 0$ and omit all the terms proportional 
to $k^2$, $k^4$, etc., in the square brackets on the r.h.s. of Eq. (\ref{Pk}). 
Therefore, this particular normalization can be only obtained 
under very specific conditions, cf. the discussion in the Appendix of
Ref.~\cite{Han1}. 

\section{Summary}

In the present paper we have studied some aspects of effects from the
final state interaction in the meson-production reaction 
$NN\rightarrow NNx$ near threshold. Specifically, 
we have demonstrated that the nucleon-nucleon FSI 
cannot be factorized from the production amplitude if one wants to obtain
reliable quantitative predictons. This conclusion confirms the 
arguments given in the paper \cite{Han}.
Furthermore, we have demonstrated that the absolute value of the 
FSI factor depends on the momentum transfer, i. e. on the mass of 
the produced meson. It is not universal! 
Only for large momentum transfers, i. e. for the production of 
heavy mesons, the FSI factor is getting independent of the mass of the 
produced meson. Finally, we have shown that the use of the Jost
function of some arbitrary potentials for the evaluation of FSI 
effects is rather questionable and may lead to a considerable 
over-estimation of those FSI effects.

\bigskip

\noindent
{\bf Acknowledgements} 

\medskip

\noindent
The authors are thankful to C. Hanhart and K. Nakayama for useful 
discussions and careful reading of the manuscript. 
A.K. acknowledges the hospitality of the Institute
f\"ur Kernphysik, Forschungzentrum J\"ulich.
A.K. and V.B. are also thankful to the Russian Foundation
for Basic Research (grant 98-02-17618) for partial support.

\vspace*{2cm}

\newpage

\end{document}